\documentclass[conference]{IEEEtran}
\usepackage{graphicx,amsmath,amsfonts,amscd,amssymb,bm,cite,epsfig,epsf,url,color}
\usepackage{fullpage} \usepackage[small,bf]{caption}
\newtheorem{theorem}{Theorem}

\newtheorem{definition}[theorem]{Definition}

\newcommand{\R}{\mathbb{R}}

\renewcommand{\>}{\rangle}

\renewcommand{\P}{\operatorname{\mathbb{P}}}
\newcommand{\E}{\operatorname{\mathbb{E}}}

\newcommand{\opnorm}[1]{\|#1\|}
\newcommand{\fronorm}[1]{\|#1\|_{F}}

\newcommand{\twonorm}[1]{\|#1\|_{\ell_2}}
\newcommand{\oneinfnorm}[1]{\|#1\|_{1,\infty}}

\newcommand{\PO}{\mathcal{P}_\Omega}

\newcommand{\cA}{\mathcal{A}}

\newcommand{\mtx}[1]{{#1}}

\newcommand{\rank}{\operatorname{rank}}

\numberwithin{equation}{section}

\title{Accurate low-rank matrix recovery from a small number of linear measurements}
\author{Emmanuel J. Cand\`es$^{\dagger}$ and Yaniv Plan$^{\sharp}$\\
  $\dagger$ Department of Statistics,
  Stanford\\
  $\sharp$ Applied and Computational Mathematics, California Institute of Technology}

\date{\today}

\begin{document}
\maketitle
\begin{abstract}
  We consider the problem of recovering a low-rank matrix $M$ from a
  small number of random linear measurements.  A popular and useful
  example of this problem is matrix completion, in which the
  measurements reveal the values of a subset of the entries, and we
  wish to fill in the missing entries (this is the famous Netflix
  problem).  When $M$ is believed to have low rank, one would ideally
  try to recover $M$ by finding the minimum-rank matrix that is
  consistent with the data; this is, however, problematic since this
  is a nonconvex problem that is, generally, intractable.

  Nuclear-norm minimization has been proposed as a tractable approach,
  and past papers have delved into the theoretical properties of
  nuclear-norm minimization algorithms, establishing conditions under
  which minimizing the nuclear norm yields the minimum rank solution.
  We review this spring of emerging literature and extend and refine
  previous theoretical results. Our focus is on providing error bounds
  when $M$ is well approximated by a low-rank matrix, and when the
  measurements are corrupted with noise.  We show that for a
  certain class of random linear measurements, nuclear-norm
  minimization provides stable recovery from a number of samples
  nearly at the theoretical lower limit, and enjoys order-optimal
  error bounds (with high probability).
\end{abstract}

\section{Introduction}
Low-rank matrix recovery is a quickly developing research area with a
growing list of applications such as collaborative filtering, machine
learning, control, remote sensing, computer vision, and quantum state
tomography.  In its most general (noiseless) form the problem consists
of recovering a low-rank matrix, $M \in \R^{n_1 \times n_2}$, from a
series of $m$ linear measurements, $\<A_1, M\>, \<A_2, M\>, ...,
\<A_m, M\>$ (we use the usual inner product $\<X, Y\> = \mbox{Tr}(X^*
Y) = \sum_{i,j} X_{i,j} Y_{i,j}$).  The $A_i$'s are known and are
analogous to the rows of a compressed sensing matrix.  To consolidate
the presentation, we write the linear model more compactly as $\cA(M)$
for the linear operator $\cA: \R^{n_1 \times n_2} \rightarrow \R^m$
(the $i$th entry of $\cA(X)$ is $\<A_i, X\>$).

If computational time were not an issue, one would ideally reconstruct
$M$ by solving
\begin{equation}
\label{eq:rank}
  \begin{array}{ll}
    \text{minimize}   & \quad \rank(X)\\
    \text{subject to} & \quad \cA(X) = \cA(M), 
 \end{array}
\end{equation}
where $X \in \R^{n_1 \times n_2}$ is the decision variable.
Unfortunately, rank minimization is an intractable problem (aside from
a few rare special cases) and is in fact provably NP-hard and hard to
approximate \cite{Grigoriev84, Meka2008}.  To overcome this problem,
nuclear-norm minimization has been introduced as the tightest convex
relaxation of rank minimization
\cite{FazelThesis,fazelrank,Recht07,CR08,CT09}. Here, one solves
instead,
\begin{equation}
\label{eq:cvx2}
  \begin{array}{ll}
    \text{minimize}   & \quad \|\mtx{X}\|_*\\
    \text{subject to} & \quad \cA(X) = \cA(M).
 \end{array}
\end{equation}
Due to its convexity, the nuclear-norm minimization problem is tractable (and an SDP) and a number of fast algorithms have been proposed to solve it \cite{SVT, Ma08}.  

A recent influx of papers has shown that for a broad range of low-rank
matrix recovery problems, nuclear-norm minimization correctly recovers
the original low-rank matrix \cite{Recht07,CT09, CR08, Ma09}.  Most of
these papers have focused on the matrix completion subproblem (see
Section \ref{sec:MC}) in which the measurements are simply entries of
the unknown matrix.  A main purpose of this paper is to compare the
theoretical results in the matrix completion problem to those possible
with `less coherent' measurement ensembles.

\subsection{Organization of the paper}
In the first half of the paper (Section \ref{sec:RIP}), we present new
theoretical results concerning low-rank matrix recovery from
measurements obeying a certain restricted isometry property, thereby extending and
refining the work of Recht et al. in \cite{Recht07}.  A first
important question we address here (and in the matrix completion
subproblem) is this: how many measurements are necessary to recover a
low-rank matrix?  By taking the singular value decomposition of $M \in
\R^{n_1 \times n_2}$ with rank $M = r$, one can see that $M$ has $(n_1
+ n_2 -r)r$ degrees of freedom.  This can be much lower than $n_1 n_2$
for $r \ll \min(n_1, n_2)$ suggesting that one may be able to recover
a low-rank matrix from substantially fewer than $n_1 n_2$
measurements.  In fact, it has been shown \cite{Recht07} that one may
oversample the degrees of freedom by a logarithmic factor and still
exactly recover $M$ via nuclear minimization with high probability.
In this paper, we show that for certain classes of linear
measurements, one can reduce the number of measurements to a small
multiple of $(n_1 + n_2 -r)r$, and still attain exact matrix recovery
via nuclear-norm minimization.  Further, when the measurements are
corrupted by noise, we suggest a nuclear norm based algorithm that
takes into account the noise in the model and show that the error when
using this algorithm is order optimal.  Lastly, when $M$ has decaying
singular values, the error bounds are refined and extended to exhibit
an optimal bias-variance trade-off (explained in more detail in
Section \ref{sec:RIP}).

In the second half of the paper (Section \ref{sec:MC}), we review the
theory on matrix completion, noting that this is a much different
problem because the RIP does not hold.  We begin the section by
comparing different theoretical results regarding nuclear norm
minimization.  We also remark that other competing algorithms have
arisen to tackle low-rank matrix completion.  To the authors' best
knowledge, only one such alternative algorithm, proposed by Montanari
et al. \cite{MontanariNoiseless, MontanariNoisy}, has rigorous
theoretical backing.  We review the theory proposed by these authors
and highlight some of the differences between their approach and
nuclear-norm minimization.  We conclude this section by reviewing the
noisy matrix completion results, and comparing them to the results
when the RIP holds.


\subsection{Notation}
In the remainder of the paper, we assume $M$ is square, with $n_1 =
n_2 = n$, in order to simplify the notation. Simple generalizations of our results,
however, hold for rectangular matrices.  
Below, $\opnorm{X}$ refers to the operator norm of $X$ (the largest
singular value), $\oneinfnorm{X}$ is the magnitude of the largest
entry of $X$
\[\oneinfnorm{X} = \max_{i,j} |X_{ij}|,\]
and $\fronorm{X}$ is the Frobenius norm.  The standard basis vectors are
denoted by $e_i$, and $\cA^*$ is the adjoint of the operator $\cA$,
$\cA:\R^{n\times n} \rightarrow \R^m$, so that
\[[\cA(X)]_i \equiv \<A_i, X\> \, \Leftrightarrow \, \cA^*(v) \equiv \sum_{i=1}^m v_i A_i.\]
The singular value decomposition of $M$ (with rank$(M) = r$) is written as
\begin{equation}
  \label{eq:svd}
  M = \sum_{i=1}^r \sigma_i u_i v_i^* =  U \Sigma V^*,
\end{equation}
with $U, V \in \R^{n \times r}, \Sigma \in \R^{r\times r}$ for
orthogonal matrices $U, V$ and the diagonal matrix of singular values,
$\Sigma$.

\section{Random linear measurements}
\label{sec:RIP}
A difficulty in the matrix completion problem is that unless all of
the entries of the unknown matrix are sampled, there is always a
rank-$1$ matrix in the null space of the sampling operator (see
Section \ref{sec:MC}).  This leads to the necessity of requirements
below on the flatness of the singular vectors of the underlying
unknown matrix. 
Interestingly, such assumptions are not necessary when considering
other classes of measurement ensembles.  In a paper bridging the gap
between compressive sensing and low-rank matrix recovery
\cite{Recht07}, the authors prove that many random measurement
ensembles often satisfy a {\em restricted isometry property} (RIP),
which guarantees that low-rank matrices cannot lie in the null space
of $\cA$ (or cannot lie `close' to the null space of $\cA$).
\begin{definition} For each integer $r = 1, 2, \ldots, n$, define the
  isometry constant $\delta_r$ of ${\cal A}$ as the smallest quantity 
  such that
  \begin{equation} \label{eq:rip}
  (1-\delta_r) \|X\|^2_F \le \|{\cal A}(X)\|_F^2 \le (1+\delta_r) \|X\|_F^2 
 \end{equation} 
holds for all matrices of rank at most $r$. 
\end{definition}
A measurement ensemble, $\cA$, is said to obey the RIP at rank  $r$ if $\delta_r \leq \delta < 1$ for a constant $\delta$ whose appropriate values will be specified in what follows. 

How many measurements, $m$, are necessary to ensure that the RIP holds
at a given rank $r$?  To first achieve a lower bound on this quantity,
note that the set of rank $r$ matrices contains the set of matrices
which are restricted to have nonzero entries only in the first $r$
rows.  This is an $n\times r$ dimensional vector space and thus we
must have $m \geq nr$ or otherwise there will be a rank-$r$ matrix in
the null space of $\cA$ regardless of what measurements are used.  The
following theorem shows that for certain classes of random
measurements, this lower bound can be achieved to within a constant
factor.

\begin{theorem} \label{teo:RIP} Fix $0 \leq \delta < 1$ and let
  $\cA$ be a random measurement ensemble obeying the following
  property: for any given $X \in \R^{n\times n}$ and any fixed
  $0<t<1$, 
\begin{equation}
\label{eq:concentration}
P(|\twonorm{\cA{X}}^2 - \fronorm{X}^2|  > t \fronorm{X}^2) \leq C \exp(-c m)
\end{equation}
for fixed constants $C,c>0$.  If $m \geq D nr$ then $\cA$ satisfies the RIP with isometry constant $\delta_r \le \delta$ with probability exceeding $1 - E e^{-d m}$ for fixed constants $D, E, d > 0$.
\end{theorem}
As an example of a generic measurement ensemble obeying \eqref{eq:concentration}, if each $A_i$ contains iid mean zero Gaussian entries
with variance $1/m$ then $m\cdot\twonorm{\cA(X)}^2/\fronorm{X}^2$
is distributed as a chi-squared random variable with $m$ degrees of
freedom.  Thus, applying a standard concentration bound,
\[P(|\twonorm{\cA{X}}^2 - \fronorm{X}^2| > t \fronorm{X}^2) \leq 2
e^{-\frac{m}{2}(\frac{t^2}{2}-\frac{t^3}{3})}
\]
and \eqref{eq:concentration} is satisfied.  Similarly, each $A_i$ can be composed of iid sub-gaussian random variables to achieve the concentration bound \eqref{eq:concentration}.  Thus one way to interpret Theorem \ref{teo:RIP} is that `most' properly normalized measurement ensembles satisfy the RIP nearly as soon as is theoretically possible, where the measure used to define `most' is Gaussian (or sub-Gaussian). 
 
Theorem \ref{teo:RIP} is inspired by a similar theorem in
\cite{Recht07}[Theorem 4.2] and refines this result in two
ways. First, it shows that one must only oversample the number of
degrees of freedom of a rank $r$ matrix by a constant factor in order
to obtain the RIP at rank $r$ (which improves on the theoretical
result in \cite{Recht07} by a factor of $\log n$).  Second, it shows
that one must only require a single concentration bound on $\cA$,
removing another assumption required in \cite{Recht07}.

\subsection{Minimax Error Bound}

Using the RIP, Recht et. al. \cite{Recht07} show that exact recovery of $M$ occurs when solving the convex problem \eqref{eq:cvx2} provided that rank($M$) $= r$ and $\delta_{5r} \leq \delta$ for a certain constant $\delta \approx .2$.  We extend this result by considering the noisy problem, 
\begin{equation}
\label{eq:noisy}
y = \cA(M) + z,
\end{equation}
where for simplicity the noise, $z$, is assumed to be Gaussian with
iid mean zero entries of variance $\sigma^2$.

In this case, we analyze the performance of a version of
\eqref{eq:cvx2} which takes noise into account, and is analogous to
the Dantzig Selector algorithm \cite{DS}:
\begin{equation}
  \label{eq:ds}
   \begin{array}{ll}
     \textrm{minimize}   & \quad \|X\|_{*}\\ 
     \textrm{subject to} & \quad \|{\cal A}^* (r)\| \le \lambda \\
     & \quad r = y - {\cal A}(X), 
  \end{array}
\end{equation}
where $\lambda = C\sqrt{n} \sigma$ for an appropriate constant $C$.  A
heuristic intuition for this choice of $\lambda$ is as follows:
suppose that $\cA$ is simply the operator which stacks the columns of
its argument into a vector, so that $\cA^* \cA$ is the identity
operator, and $\cA^*(z)$ is an $n \times n$ matrix with iid Gaussian
entries.  This is perhaps the simplest case to analyze.  We would like
the unknown matrix $M$ to be a feasible point, which requires that
$\opnorm{\cA^*(z)} \leq \lambda$.  It is well known that the top
singular value of a square $n \times n$ Gaussian matrix, with
per-entry variance $\sigma^2$, is concentrated around $\sqrt{2n}
\sigma$, and thus we require $\lambda \geq \sqrt{2n} \sigma$. Further,
observe that in this simple setting the solution to \eqref{eq:ds} can
be explicitly calculated, and is equal to $T_\lambda(M + \cA^*(z))$
where the operator $T_\lambda$ soft-thresholds the singular values of
its argument by $\lambda$.  If $\lambda$ is too large, then
$T_\lambda(M + \cA^*(z))$ becomes strongly biased towards zero, and
thus (loosely) $\lambda$ should be as small as possible while still
allowing $M$ to be feasible, leading to the choice $\lambda \approx
\sqrt{2n} \sigma$ for this simple case.

We are now prepared to present the simplest version of our theoretical
error bounds.  The following theorem states that if $M$ has low rank then the error is order optimal with overwhelming probability.
\begin{theorem} \label{teo:RIPbound}  Suppose that $\mathcal{A}$ has RIP constant $\delta_{4r} < \sqrt{2} - 1$ and rank($M$)$ = r$.  Let $\hat{M}$ be the solution to \eqref{eq:ds}.  Then
\begin{equation}
\label{eq:errorBound1}
\fronorm{\hat{M} - M}^2 \leq C n r \sigma^2
\end{equation}
with probability at least $1 - De^{-dn}$ for fixed numerical constants
$C, D, d >0$.
\end{theorem}
The result in this theorem is quite similar to the adaptive error bound in compressive sensing first proved in \cite{DS} and the proofs are almost identical (see \cite{CP10} for a proof).  In order to see how the result generalizes when $M$ is rectangular, in the case when $M \in \R^{n_1 \times n_2}$, the error bound \eqref{eq:errorBound1} is replaced by 
\[\fronorm{\hat{M} - M}^2 \leq C \max(n_1, n_2) r \sigma^2.\]

We compare the above error bound \eqref{eq:errorBound1}, to the minimax error bound described below,
\begin{theorem}
\label{teo:minimax}
  Any estimator $\hat{M}(y)$, with $y = \cA(M) +z$, obeys
  \begin{equation}
    \label{eq:minimax}
    \sup_{M : \rank(M) \le r} \, \E \|\hat M - M\|^2 \ge \frac{1}{1+\delta_r} \, n r \sigma^2.    
  \end{equation}
  In other words, the minimax error over the class of matrices of rank
  at most $r$ is lower bounded by just about $n r \sigma^2$.
\end{theorem}
Thus the error achieved by solving a convex program is within a
constant of the expected minimax error (with high probability).  As an exercise, and to help further
understand the error bound \eqref{eq:errorBound1}, we analyze the
error in the example above in which $\cA^*\cA$ is the identity
operator and $\hat{M} = T_\lambda(M + \cA^*(z))$.  In this case,
letting $\tilde{M} = M + \cA^*(z)$,
\begin{align*}
\opnorm{\hat{M} - M} &= \opnorm{T_\lambda(\tilde{M}) - \tilde{M} + \cA^*(z)}\\
&\leq \opnorm{T_\lambda(\tilde{M}) - \tilde{M}} + \opnorm{\cA^*(z)}\\
&\leq 2\lambda
\end{align*}
assuming that $\lambda \geq \opnorm{\cA^*(z)}$.  Then, 
\begin{align}
  \nonumber \fronorm{\hat{M} - M}^2 & \leq \opnorm{\hat{M} - M}^2 ~
  \mbox{rank}(\hat{M} - M) \\
& \leq 4 \lambda^2 ~ \mbox{rank}(\hat{M} -
  M).
\end{align}
Once again, assuming that $\lambda \geq \opnorm{\cA^*(z)}$, we have
$\mbox{rank}(\hat{M} - M) \leq \mbox{rank}(\hat{M}) + \mbox{rank}(M)
\leq 2r$.  Plugging this in with $\lambda = C \sqrt{n} \sigma$ gives
the error bound \eqref{eq:errorBound1}.

\subsection{Oracle Error Bound}

While achieving the minimax error is useful, in many cases minimax analysis is overly focused on worst-case-scenarios and more adaptive error bounds can be reached.  This is exactly the case when $M$ has decaying singular values, with many singular values below the `noise level' of $\sqrt{n} \sigma$.  In order to set the bar for error bounds in this case, we compare to the error achievable with the aid of an oracle.

To develop an oracle bound, consider the family of estimators defined
as follows: for each $n \times r$, orthogonal, matrix $U$, define $\hat{M}[U]$ as the minimizer to \eqref{eq:MU}
\begin{equation}
  \label{eq:MU}
 \min \{\|y - {\cal A}(\hat M)\|_{\ell_2} : 
\hat M = UR \text{ for some } R\}.   
\end{equation}
In other words, we fix the column space (the linear space spanned by
the columns of the matrix $U$), and then find the matrix with that
column space which best fits the data. Knowing the true matrix $M$, an
oracle or a genie would then select the best column space to use as to
minimize the mean-squared error (MSE)
\begin{equation}
  \label{eq:oracle}
 \inf_U \E \|M - \hat{M}[U]\|^2. 
\end{equation}
The question is whether it is possible to mimic the performance of the
oracle and achieve a MSE close to \eqref{eq:oracle} with a real
estimator.

Through classical calculations, one may lower bound $\|M -
\hat{M}[U]\|^2$ (the steps required will be shown in detail in the
sequel) as follows: we have 
\[
\E \|M - \hat{M}[U]\|_F^2 \ge \, \left[
\|P_{U^\perp}(M)\|_F^2 + \frac{nr \sigma^2}{1+\delta_r} \right], 
\]
where $P_{U^\perp}(M) = (I - UU^*) M$.  The first term is a bound on
the bias of the estimator which occurs when $U$ does not span the
column space of $M$ while the second term is a bound on the variance
which grows as the dimension of $U$ grows.  Thus the oracle error is
lower bounded by 
\[
\inf_U \E \|M - \hat{M}[U]\|_F^2 \ge \inf_U \, \left[
\|P_{U^\perp}(M)\|_F^2 + \frac{nr \sigma^2}{1+\delta_r} \right]. 
\]
Now for a given dimension $r$, the best $U$---that minimizing the
proxy for the bias term $\|P_{U^\perp}(M)\|_F^2$---spans the
top $r$ singular vectors of the matrix $M$ and thus we obtain
\[
\inf_U \E \|M - \hat{M}[U]\|_F^2 \ge \inf_r \, \left[\sum_{i > r}
  \sigma_i^2(M)
  \ + \frac{1}{2} nr \sigma^2\right],
\]
which for convenience we simplify to
\begin{equation}
  \label{eq:oraclebound}
  \inf_U \E \|M - \hat{M}[U]\|_F^2 \ge \frac{1}{2} 
\sum_i \min(\sigma_i^2, n\sigma^2). 
\end{equation}
The right-hand side has a nice interpretation. If $\sigma_i^2 > n \sigma^2$, one should try to estimate the
rank-$1$ contribution $\sigma_i u_i v_i^*$ and pay the variance term
(which is about $n\sigma^2$) whereas if $\sigma_i^2 \le n \sigma^2$,
we should not try to estimate this component, and pay a squared bias
term equal to $\sigma_i^2$. In other words, the right-hand side may be
interpreted as an ideal bias-variance trade-off, which can be nearly
achieved with the help of an oracle. 

The following theorem states that when $M$ has low rank, one achieves
the optimal bias-variance trade-off when solving a convex optimization
problem, up to a constant factor.
 
\begin{theorem} \label{teo:oracleBound} Suppose that $\mathcal{A}$ has
  RIP constant $\delta_{4r} < \sqrt{2} - 1$ and rank($M$)$=r$.  Let
  $\hat{M}$ be the solution to $\eqref{eq:ds}$.  Then
\[\fronorm{\hat{M} - M}^2 \leq C \sum_{i=1}^r \min(\sigma_i^2, n\sigma^2)\]
with probability at least $1 - De^{-dm}$ for some numerical constants
$C,D, d > 0$.
\end{theorem}
For a proof, see the upcoming paper \cite{CP10}.
\subsection{Approximately low-rank, noisy, error bounds}

An important drawback of the above two theorems (Theorems \ref{teo:oracleBound}, \ref{teo:RIPbound}) is that they only apply when $M$ is exactly a low-rank matrix, 
but do not generally apply when $M$ is well approximated by a low-rank matrix.
However, for
many random measurement ensembles $\cA$, the above result can be
extended to handle the case when all $n$ of the singular values of $M$
are nonzero.  This is the content of the following theorem.
\begin{theorem} \label{teo:instanceOpt} Fix $M$.  Suppose that each
  `row' $A_i$ of $\cA$ contains iid mean zero Gaussian entries with variance $1/m$. Suppose $m \leq c
  n^2/\log n$ for some numerical constant $c$.  Let $\bar{r}$ be the
  largest integer such that $\delta_{4r} \leq \frac{1}{2} (\sqrt{2} -
  1)$.  Let $\hat{M}$ be the solution to $\eqref{eq:ds}$. Then
\begin{equation}
\fronorm{\hat{M} - M}^2 \leq C\left(\sum_{i=1}^{\bar{r}} \min(\sigma_i^2, n\sigma^2) + \sum_{i = \bar{r}+1}^n \sigma_i^2 \right)
\end{equation}
with probability greater than $1 - De^{-dn}$ for fixed numerical
constants $C,D, d> 0$.
\end{theorem}
Here, $\bar{r}$ is the largest value of $r$ such that the RIP holds and
thus $\bar{r} \geq c\frac{m}{n}$ with high probability for a fixed numerical
constant $c$ (see Theorem \ref{teo:RIP}).  The constant $\frac{1}{2}$
in $\delta_{4r} \leq \frac{1}{2} (\sqrt{2} - 1)$ is arbitrary and
could be replaced by any constant less than 1.  The error bound has an
interesting intuitive interpretation: decompose $M$ as $M = M_{\bar{r}} +
M_c$ with
\[M_{\bar{r}} = \sum_{i=1}^{\bar{r}} \sigma_i u_i v_i^*, \quad M_c =
\sum_{i>\bar{r}} \sigma_i u_i v_i^*\] so that $M_{\bar{r}}$ is the
projection of $M$ onto rank-$\bar{r}$ matrices.  Then we achieve the
near optimal bias-variance trade-off in estimating $M_{\bar{r}}$, but
cannot recover $M_c$.

An important point about Theorem \ref{teo:instanceOpt} is that it is
an example of instance optimality: the result holds with high
probability for any given specific $M$, but it does not hold uniformly
over all $M$.  For the proof, see \cite{CP10}.

\section{Matrix completion}
\label{sec:MC}
A highly applicable subset of low-rank matrix recovery problems
concerns the recovery of an unknown matrix from a subset of its
entries (matrix completion).  An example to bear in mind is the
Netflix problem in which one sees a few movie ratings for each user,
which can be viewed as a row of (possible) ratings with only a few
entries filled in.  Stacking the rows together, creates the data matrix. Netflix would like to guess how each user would rate a movie
he had not seen, in order to target advertising.
 A great difficulty is that there are always rank-1 matrices in the
null space of the measurement operator and, thus, our problem is
`RIPless'.

In order to specialize the nuclear-norm minimization algorithm
\eqref{eq:cvx2} to matrix completion, let $\Omega$ be the set of
observed entries.  We assume $\Omega$ is chosen uniformly at random
with $|\Omega| = m$ (this turns the discussion away from adversarial
sampling sets).  Define $P_\Omega: \R^{n\times n} \rightarrow
\R^{n\times n}$ to be the operator setting to zero each unobserved
entry,
\begin{equation}
[P_\Omega(X)]_{ij} = \left\{\begin{tabular}{ll}
$X_{ij}$,&$\quad \mbox{if } (i,j) \in \Omega$\\
$0 $,&$\quad \mbox{if } (i,j) \notin \Omega$.\\
\end{tabular}\right.
\end{equation}
Then one solves 
\begin{equation}
\label{eq:mc}
  \begin{array}{ll}
    \text{minimize}   & \quad \|\mtx{X}\|_*\\
    \text{subject to} & \quad P_\Omega(X) = \P_\Omega(M).
 \end{array}
\end{equation}

To the best of our knowledge, there are five papers with novel
theoretical guarantees on noiseless matrix completion \cite{CR08,
  CT09, Ma09, MontanariNoiseless, MontanariNoisy}.  We compare the
results of this prior literature in Table \ref{tab:comparisons}.  The
parameters $\mu, \mu_1, \mu_2, \mu_B, \kappa$ in Table
\ref{tab:comparisons} are defined further on in this section, but for
now note that they depend on the structure of the underlying matrix,
$M$, and in many cases are small (e.g. $O(1)$ or $O(\log n)$) under
differing assumptions on $M$.  

\begin{table}[t]
\scriptsize
\begin{center}
\begin{tabular}{|c|c|c|}
\hline
Assumptions        & Number of measurements                         & Paper/\\
        on $M$     & $m$ required                                   & Theorem\\
\hline
$M$ is    & $Cn^{5/4}r \log(n)$                            &\cite{CR08},\\
generic * & or                                             & Thm 1.1\\
              & $Cn^{6/5}r \log(n)\, \mbox{ if } r\leq n^{1/5}$  &\\
\hline
none               & $C\max(\mu_1^2, \mu_0^{1/2} \mu_1, \mu_0 n^{1/4})nr \log n $   &\cite{CR08},\\
                   & or                                             & Thm 1.3\\
                   & $C \mu_0 n^{6/5}r \log(n) \,\mbox{ if } r\leq \mu_0^{-1} n^{1/5}$ &\\
\hline    
$M$ is  & $C nr \log^8 n$                                & \cite{CT09},\\
generic *  & or                                             & Cor. 1.6\\
              & $C nr \log^7 n \,\mbox{ if } r \geq \log n$      &\\
                   & $C nr \log^6 n \,\mbox{ if } r = O(1)$           &\\
\hline
$r = O(1)$         & $C \mu_B^4 n \log^2 n$                         & \cite{CT09},\\
                   &                                                & Cor. 1.5\\
\hline

none               & $C \mu^2 nr \log^6 n$                          & \cite{CT09},\\
                   &                                                & Thm 1.2\\
\hline
$M$ is  & $\max(c_2 n^2, m_0)$ **                                     & \cite{Ma09},\\
generic *, &                                                & Thm 2.5\\
 $r \leq c_1 n$\hspace{2mm} **&                                           &\\
\hline
none               & $C n \kappa^2 \max(\mu_0 r \log n, \mu_0^2 r^2 \kappa^2, \mu_2^2 r^2 \kappa^4)$ & \cite{MontanariNoisy},\\
                   &                                                & Thm 1.2\\
\hline 
\end{tabular}
\end{center}
\caption{Comparison of different theoretical guarantees for matrix completion.  When the requirements on $M$ and the number of measurements are met, and the measurements are chosen uniformly at random, then exact matrix completion is guaranteed with probability at least $1 - cn^{-3}$ (for a fixed constant $c$).  $C$ is also a fixed constant.  The algorithm used to produce the results in the last line is OPTSPACE, the rest of the table refers to nuclear-norm minimization \eqref{eq:mc}.  \newline * $M$ is drawn from the random orthogonal model which is defined below.  Intuitively, under this model the singular vectors of $M$ have no structure and are thus `generic'. \newline ** The constants $c_1$ and $c_2$ satisfy $c_1, c_2 < 1$ and $m_0$ is a fixed integer.}
\label{tab:comparisons}
\end{table}

\subsection{nuclear-norm minimization algorithms}
We first review the results of \cite{CR08}, which pioneered the matrix
completion theory.  As described therein, assumptions on $M$ are vital
to ensure that matrix completion is possible.  To compel this line of
reasoning, suppose $M = e_i e_j^*$ is a (rank-1) matrix with only 1
nonzero entry.  If this entry is not seen, then $M$ is in the null
space of the measurement operator and is indistinguishable from the
zero matrix.  Such observations are explored in more depth in
\cite{CP09, CR08, CT09} providing an argument for the necessity of the
assumption that the singular vectors of $M$ are `spread', which is
also intrinstically important to bounding the size of $\mu_B, \mu_0,
\mu_1, \mu_2$ and $\mu$ (but has no relation to $\kappa$).

In order to quantify `spread', with parameter $\mu_B$, the authors of \cite{CR08} require
\begin{equation}
  \label{eq:bdd}
  \|u_k\|_{\ell_\infty}, \|v_k\|_{\ell_\infty} \le \sqrt{\mu_B/n}, 
\end{equation}
for each $u_k, v_k$ (recall these are the singular vectors of $M$).  Note that the minimum value of $\mu_B$ is 1 if all of the singular vectors have minimal $\ell_\infty$ norm, and that $\mu_B$ can be as large as $n$ when a singular vector has only one nonzero entry.  When $r = O(1)$, the constants $\mu_0, \mu_1$ and $\mu$ are all $O(1) \cdot \mu_B$ (see \cite{CT09, CR08}), thus bounding all of the parameters involved in the nuclear norm theoretical results. 

In order to prove theoretical guarantees for larger values of the
rank, \cite{CR08} introduces the concept of the incoherence of $M$
with parameters $\mu_0$ and $\mu_1$ as defined below.  Let $P_U =
UU^*$ be the projection onto the range of the left singular vectors of
$M$ and similarly let $P_V = VV^*$.  Then \cite{CR08} requires,
\begin{align*}
\max_{1 \leq i \leq n} \twonorm{P_U e_i}, \max_{1 \leq i \leq n} \twonorm{P_v e_i} &\leq \sqrt{\frac{r}{n}} \mu_0,\\
\oneinfnorm{UV^*}&\leq \frac{\sqrt{r}}{n} \mu_1. 
\end{align*}
A matrix $M$ is said to be incoherent if $\mu_0$ and $\mu_1$ are small (e.g. $O(1)$ or $O(\log n)$...).  Note that these parameters, and thus the number of measurements required in Theorem 1.3 of \cite{CR08} have no dependence on the singular values of $M$, a quality that is ubiquitous to all of the parameters involved in the nuclear-norm minimization theory.

Which matrices are incoherent?  As noted above, if $r = O(1)$ then
$\mu_0, \mu_1 \leq O(1)\cdot \mu_B$ and thus the matrices with
`spread' singular vectors are incoherent.  To address this question
from another angle, introduce the random orthogonal model mentioned in
Table \ref{tab:comparisons}.
\begin{definition}
A matrix $M = U \Sigma V^*$ of rank $r$ is said to be drawn from the random orthogonal model if $U$ is drawn uniformly at random from the set of $n \times r$ orthogonal matrices and similarly for $V$, although $U$ and $V$ may be dependent on each other.
\end{definition}
This is perhaps the most generic possible random model for the singular vectors of a matrix.  Under this model for values of the rank $r$ greater than $\log n$ (to avoid small sample effects) $\mu_0 = O(1)$ and $\mu_1 = O(\log n)$ with very large probability \cite{CR08}.  A way to interpret this is that `most' matrices have small values of $\mu_0, \mu_1$.  

With the variables $\mu_0$ and $\mu_1$ defined, along with the random
orthogonal model, the reader is equipped to evaluate the theoretical
results of \cite{CR08} in Table \ref{tab:comparisons}.  One sees that
for `most' matrices, or alternatively, for incoherent matrices (those
with small values of $\mu_0, \mu_1$), it is required that $m \gtrsim
n^{1.2} r$ or $m \gtrsim n^{1.25} r$ (depending on $r$), ignoring
$\log$ and constant factors.  While these results show that one can
drastically undersample a matrix when $r \ll n$, they are above the
theoretical limit of $(2n - r)r \approx nr$ by a factor of about
$n^{.2}$ or $n^{.25}$.  With the aid of some slightly stronger
assumptions on $M$, \cite{CT09} removes these extra small
powers of $n$ and nearly attains the theoretical limit.

In order to present these optimal results \cite{CT09} that apply for
values of the rank $r$ greater than $O(1)$, the authors introduce the
{\em strong incoherence property} with parameter $\mu$, which we now state:
it is required that for all pairs $(a,a')$ and $(b,b')$ with $1\leq
a,a',b,b'\leq n$,
\begin{align*}
\Bigl|\<e_a, P_U e_{a'}\> - \frac{r}{n} 1_{a = a'}\Bigr| & \le \mu \frac{\sqrt{r}}{n},\\
\Bigl|\<e_b, P_V e_{b'}\> - \frac{r}{n} 1_{b = b'}\Bigr| & \le \mu \frac{\sqrt{r}}{n}.
\end{align*}
Secondly, it is required that $\mu \geq \mu_1$ (with $\mu_1$ defined
above).  As in \cite{CR08}, the random orthogonal model obeys $\mu
\leq O(\log n)$ with high probability
\cite{CT09}.  
Examining Table \ref{tab:comparisons}, one sees that for $\mu = O(\log
n)$, the number of measurements required is within a polylogarithmic
factor of the theoretical low limit.

Is the polylogarithmic factor necessary in the bounds above?  This answer depends on the size of $r$.  As argued in \cite{CR08}, \cite[Theorem 1.7]{CT09}, when $r= O(1)$ it is generally impossible to recover $M$ by any algorithm if one does not oversample the degrees of freedom by at least a factor of $\log n$.  However, as shown in \cite{Ma09}, when $r$ is of the same order as $n$ and $M$ is drawn from the random orthogonal model, one can oversample the degrees of freedom by a constant factor (while still undersampling $M$), and still have exact recovery with high probability.

\subsection{OPTSPACE}
We now turn to the algorithm OPTSPACE proposed in
\cite{MontanariNoiseless, MontanariNoisy}.  This algorithm has three
steps, as (roughly) described below.
\begin{itemize}
\item[(1)] Remove the columns and rows that contain a disproportionate amount of sampled entries (trimming) in order to prevent these measurements from overly influencing the singular vectors in the next step.
\item[(2)] Project the result of step 1 onto the space of rank $r$ matrices and renormalize in order to attain an initial approximation of $M$. \footnote{It is assumed that $r$ is known in this step.  The authors of \cite{MontanariNoiseless, MontanariNoisy} suggest to estimate $r$ using the trimmed matrix from step 1, or to test different values of $r$.}
\item[(3)] Perform local minimization via gradient descent over a
  locally convex, but globally nonconvex, function $F(\cdot)$
  described in \cite{MontanariNoiseless, MontanariNoisy}, which has
  $M$ as a local minimum.
\end{itemize}
The intuitive idea of the algorithm is that the first 2 steps provide
an accurate initial guess for $M$ and that the function $F(\cdot)$
behaves like a parabola near $M$ (with $M$ achieving the minimum of
the parabola) and thus gradient descent will recover $M$.

The success of OPTSPACE is theoretically tied to the values of the
parameters $\kappa, \mu_0$ and $\mu_2$.  The last has been introduced
while the first is the condition number
\[
\kappa \equiv \sigma_1/\sigma_r.
\]
The parameter $\mu_2$ is somewhat
analogous to $\mu_1$ above.  In fact, \cite{MontanariNoisy,
  MontanariNoiseless} require
\[
\oneinfnorm{\sum_{i=1}^r \frac{\sigma_i}{\sigma_r} u_i v_i^*} \leq \frac{\sqrt{r}}{n} \mu_2
\]
In the special case where the singular values of $M$ are all equal so
that $\kappa = 1$, $\mu_1$ and $\mu_2$ have equivalent definitions,
compelling the intuition that when $\kappa = O(1)$ the two parameters
are comparable.  In this setting, and if $r = O(\log n)$,
\cite{MontanariNoisy} poses strong theoretical results, comparable to
those of \cite{CT09}, but with smaller powers of the parameters
involved and the logarithms.  However, the applicability of the theory
depends strongly on the assumption that $\kappa$ is small, whereas
when using nuclear-norm minimization, the variations in the nonzero
singular values are inconsequential to the exact recovery results.

\subsection{Noisy matrix completion}

As explained above, there is always a rank-1 matrix in the null space
of the operator sampling the entries, and thus the RIP does not hold.
To understand the difficulty this creates, consider that in the
related field of compressive sensing, `RIPless' error bounds have
proved extremely elusive.  To the authors' best knowledge, there is
only one paper with such results \cite{CP07}, but it requires that
every element of the signal should stand above the noise level.
Despite this difficulty, two recent papers \cite{CP09, MontanariNoisy}
prove that matrix completion is robust vis-a-vis noise (using
nuclear-norm minimization in \cite{CP09} and OPTSPACE in
\cite{MontanariNoisy}).  In order to state these results, we first
specify the noisy matrix completion problem.

The noisy model assumes
\begin{equation}
  Y_{ij} = M_{ij} + Z_{ij}, \quad (i,j) \in \Omega,  
\end{equation}
where $\{Z_{ij} : (i,j) \in \Omega\}$ is a noise term and, as before,
$\Omega$ is chosen uniformly at random with $|\Omega| = m$. Another
way to express this model is as
\[
\PO(Y) = \PO(M) + \PO(Z), 
\]
for some noise matrix $Z$ (the entries of $Z$ outside of $\Omega$ are
irrelevant).

\subsection{Stability with nuclear-norm minimization}

The recovery algorithm analyzed in \cite{CP09} is a relative of the
Dantzig Selector, and once again draws its roots from an analogous
algorithm in compressive sensing, this time the Lasso:
\begin{equation}
  \label{eq:lasso}
   \begin{array}{ll}
     \textrm{minimize}   & \quad \|X\|_{*}\\ 
     \textrm{subject to} & \quad \twonorm{\PO(X) - \PO(M)} \le \delta.
  \end{array}
\end{equation}
This time, $\delta$ should be larger than the Frobenius norm of the
noise, i.e. $\delta \geq \fronorm{\PO(Z)}$---at least
stochastically.\footnote{For example, if the entries of $Z$ are iid
  $N(0, \sigma^2)$, one may take $\delta^2 = (m +
  \sqrt{8m})\sigma^2$.}  Thus, the algorithm just minimizes the proxy
for the rank, while keeping within the noise
level. 

The claim in \cite{CP09} is that as soon as noiseless matrix
completion is possible via nuclear-norm minimization, so is stable
matrix completion (this argument is made in detail in \cite{CP09}).
We distill this result into the following simple theorem:
\begin{theorem} \cite{CP09}
\label{teo:stable}
Suppose that any of the requirements in \cite{CR08} or \cite{CT09} for
exact matrix completion in the noiseless case are met (see Table
\ref{tab:comparisons}).  Suppose $\fronorm{\PO(Z)} \leq \delta$.  Let $p = m/n^2$.  Then
the solution to \eqref{eq:lasso}, $\hat M$, obeys
  \begin{equation}
    \label{eq:stable}
    \|\hat{M} - M\|_F \le  4 \sqrt{\frac{C_p n}{p}} \, \delta + 2  \delta, \quad C_p = 2 + p,
  \end{equation}
  with probability at least $1 - cn^{-3}$ for a fixed numerical
  constant $c$.
\end{theorem}

While this result is noteworthy in that it has no current analogue in
compressive sensing\footnote{The authors are in the process of writing
  an analogous paper for the compressive sensing case.}, it falls
short of achieving oracle type error bounds.  As described in
\cite{CP09} an oracle error bound derived by giving away the column
space of $M$ in the noisy matrix completion problem is
\[\fronorm{M^{\text{Oracle}} - M} \approx p^{-1/2} \delta\]
(this oracle error is focused on adversarial noise).  One sees that the oracle error is over-estimated by a factor of about $\sqrt{n}$. 

\subsection{Stability with OPTSPACE}
Another recent and noteworthy theoretical error bound for noisy matrix
completion appears in a paper by Montanari et
al. \cite{MontanariNoisy}.  Once again the OPTSPACE algorithm is used,
and thus having a large spread in the singular values of $M$ can cause
instabilities.  However, as described in the following theorem, under
suitable conditions the error bounds are comparable to those
achievable with the aid of an oracle (with stochastic noise).
\begin{theorem} \cite{MontanariNoisy}
Suppose rank$(M) = r$ and 
\[ m \geq C n \kappa^2 \max(\mu_0 r \log n, \mu_0^2 r^2 \kappa^2,
\mu_2^2 r^2 \kappa^4)\] for a fixed numerical constant $C$.  Let
$\hat{M}$ be the solution to the OPTSPACE algorithm.  Then
\[\fronorm{\hat{M} - M} \leq C^\prime \kappa^2 \frac{n^2\sqrt{r}}{m} \opnorm{\PO(Z)} \]
with probability at least $1-1/n^3$, assuming that the RHS is smaller
than $\sigma_r$, for a fixed numerical constant $C^\prime$.
\end{theorem}
Here $\sigma_r$ is the smallest nonzero singular value of $M$.

When
$Z$ contains iid Gaussian entries with variance $\sigma^2$, the term
$\opnorm{\PO(Z)}$ can be bounded as
\[
\opnorm{\PO(Z)} \leq C\left(\frac{m \log n}{n}\right)^{1/2} \sigma
\]
with high probability (see \cite{MontanariNoisy}).  Thus, in the
regime when $\kappa = O(1)$ and $\sigma_r \geq C^\prime \kappa^2
\frac{n^2\sqrt{r}}{m} \opnorm{\PO(Z)}$, one has
\[\fronorm{\hat{M} - M}^2 \leq C \frac{n^3 r \log n}{m}  \sigma^2\]
which is within a logarithmic factor of a simple oracle bound discussed in
\cite{CP09}, in which the exact column space is given away and the noise is assumed to be stochastic.  Specifically, this is the oracle bound that one achieves by examining the expected error of the estimator $\hat{M}[U]$ defined in equation \eqref{eq:MU}, where $U$ is defined as in the SVD $M = U\Sigma V^*$.

However, the class of low-rank matrices to which the theorem applies
is very restrictive, a problem that is non-existent when the RIP
holds.  In order to see this, note first that it is required that all of the singular values of $M$ stand far above the noise level.  For example, if one sees the entire matrix ($m=n^2$) then the theorem requires $\sigma_r \geq C^\prime \kappa^2\sqrt{r} \opnorm{Z}$, i.e. the minimal singular value of $M$ must be larger than the noise level by a factor of about $\kappa^2 \sqrt{r}$.  Secondly, the number of measurements required is at least $C \kappa^6 \mu_2^2 r^2$ and thus quickly grows much larger than the degrees of freedom of $M$ when $\kappa$ and $r$ grow.


\section{Conclusion} 
We have shown that a nuclear-norm minimization algorithm \eqref{eq:ds}
recovers a low-rank matrix from the noisy data $\<A_i, M\> + z_i$, $i
= 1, \ldots, m$, in which each $A_i$ is Gaussian (or sub-Gaussian),
and enjoys the following properties:
\begin{enumerate}
\item For both exact recovery from noiseless data and accurate
  recovery from noisy data, the number of measurements $m$ must only
  exceed the number of degrees of freedom by a constant factor.
\item With high probability the error bound is within a constant factor of the expected minimax error.
\item With high probability the error bound achieves an optimal bias-variance trade-off (up
  to a constant).
\item The error bounds extend to the case when $M$ has full rank (with
  many `small' singular values).
\end{enumerate}

We close this paper with a few questions that we leave open for future
research. Can the `RIPless' theoretical guarantees be improved? In
particular, in the case of nuclear-norm minimization based algorithms,
can the error bound be tightened? And for other tractable algorithms,
can we achieve strong error bounds without requiring the nonzero
singular values of $M$ to be nearly constant? Finally, are there
useful applications in which the measurements are `incoherent'
enough that the RIP provably holds?

\bibliographystyle{plain}
\bibliography{ref}
\end{document}